UseRawInputEncoding
\documentclass
[prl,aps,onecolumn,showpacs]{revtex4}

\usepackage{psfrag}
\usepackage{graphicx}
\usepackage{graphics}
\usepackage{bm}
\usepackage{color}
\usepackage{centernot}

\newcommand{\be}{\begin{equation}}
\newcommand{\ee}{\end{equation}}
\newcommand{\ba}{\begin{eqnarray}}
\newcommand{\ea}{\end{eqnarray}}

\newcommand{\dcom}[1]{}
\newcommand{\dnote}[1]{}

\newcommand{\gsim}{\raise.3ex\hbox{$>$\kern-.75em\lower1ex\hbox{$\sim$}}}
\newcommand{\lsim}{\raise.3ex\hbox{$<$\kern-.75em\lower1ex\hbox{$\sim$}}}

\begin{document}

\renewcommand{\thefootnote}{\fnsymbol{footnote}}


\renewcommand{\thefootnote}{\arabic{footnote}}
\setcounter{footnote}{0} \typeout{--- Main Text Start ---}

\title{ Variable Modified Newtonian Mechanics IV: Non Rotating Galaxies}
\author{ James C.~ C. Wong}
\affiliation{Department of Electrical and Electronic Engineering, University
of Hong Kong. H.K.}

\date{\today}
\begin{abstract}
At it stands, the $\Lambda CDM$ model does not anticipate the early emergence of massive galaxies \cite{mcgaugh2024}. Canonical Modified Newtonian Dynamics (MOND) faces difficulties at late time solar system scale \cite{tremaine}-\cite{desmond} and Wide-Binary scales \cite{banik}. To match data, a MOND variant needs a variable MOND acceleration $a_0$ which is strong at high redshift galactic scale and diminishes over redshift to far below Newtonian gravity at solar system scale at late time. We found such a candidate in a relativistic frame-work. In \cite{wong}, a new single-metric solution of Einstein Gravity is found for a point mass residing in an expanding universe, which apart from the Newtonian acceleration, gives rise to an additional MOND-like acceleration in which the MOND acceleration $a_0$ is replaced by the cosmological acceleration $\frac{1}{2}H^2(z)r$. This cosmological acceleration is shown to be far below Newtonian acceleration in the late time solar system and therefore avoids the problem of MOND program.
\\\\
In this work, we study the monolithic evolution of a $10^{10.5}M_{\odot}$ (Milky Way mass) protogalactic cloud at recombination in this model where the non-Newtonian acceleration is stronger than Newtonian gravity. To obtain a spherical galaxy we assume that a point on a mass shell at turnaround will pick up sufficient non-systematic angular momentum. Assuming a violent relaxation process similar to the simulation studies for Newtonian gravity and MOND, we find that the central core can form a time independent Quasi-Stationary-State (QSS) by $z\gsim 7$, which could explain the galaxy morphology stability observations for $z<6.5$ in \cite{ferreira2}. The virialised potential has a Newtonian acceleration dominant central region and a MOND-like acceleration dominant outer region. We evaluate the corresponding MOND acceleration $a_0^{VM}$ in a virialised potential for a $10^{10.5}M_{\odot}$ elliptical galaxy and find that $a_0^{VM}\sim a_0$.
\end{abstract}

\pacs{??}

\maketitle
\section{Introduction}
Mass discrepancy problems in astrophysics and cosmology are modelled extensively by working within General Relativity (GR) with an introduction of an appropriate amount of postulated cold dark matter (CDM) particles in the Stress Momentum Tensor. At very large scale, the presence of Dark matter potential after radiation-matter equality provides a good explanation for the Cosmic Microwave Background (CMB) angular power spectrum \cite{mukhanov} and constrains the dark matter density parameter as well as the present time Hubble constant \cite{planck}. The constrained cosmological parameters specify the standard $\Lambda CDM$ model. At the matter power spectrum scale, the modelled $\sigma_8$ at small redshift observation begins to develop tension \cite{bohringer}-\cite{einasto}. At late time galactic scale, there are modelling difficulties \cite{vand}-\cite{mcgaugh2018}. In fact, observations support an unknown correlation between dark matter halo and its baryonic distribution \cite{salucci}. At late time solar system scale, dark matter halo density is low and observation is consistent with Baryonic matter dominance, but the observed Hubble constant is in tension with the constrained value from CMB power spectrum \cite{riess}-\cite{valentino}. The high redshift observation of the James Webb Space Telescope (JWST) provides the biggest challenge to the $\Lambda CDM$ paradigm, which does not anticipate the early emergence of both massive galaxies \cite{mcgaugh2024} and Supermassive Black Holes (SMBH) \cite{cho}. The formation of these objects seems to require a significant increase of gravity soon after recombination beyond the standard $\Lambda CDM$ model provides. As it stands, the postulated Dark matter particles remain to be found experimentally and the theoretically viable candidates are getting fewer.
\\\\
An alternative approach is to search for new non-Newtonian force law (or gravitational potential) without the need of adding CDM. The simplest such paradigm is MOND proposed by Milgrom \cite{milgrom} with its inclusion of a non-Newtonian radial acceleration, for a central mass $M$,
\begin{equation}
\ddot{r}=-\sqrt{g_N a_0}, \:\:\: g_N=\frac{GM}{r^2},
\label{mond}
\end{equation}
which dominates the gravitational acceleration at Newtonian acceleration $g_N$ where $g_N\ll a_0$.  In this paradigm, the value of the acceleration scale $a_0$ is taken to be an universal constant. 
Alternatively, for a given averaged galactic rotation speed $V$ around a central mass $M$, one could work from the phenomenological Tully-Fisher relation $V^4=GMa_0$ (with $\ddot{r}=-V^2/r$) and arrives at a heuristic model with the MOND acceleration equation Eq.(\ref{mond}), where the acceleration scale $a_0$ is not a priori fixed but is given by observations of the system under consideration. It can either be an Universal constant or be dependent on the parameters of the system under consideration. 
\\\\
The small observed scatter of the empirical BTFR of gas rich spiral galaxies at the scale $a_0=1.2\times 10^{-10}ms^{-2}$, McGaugh \cite{mcgaugh2004}-Lelli \cite{lelli}, makes a strong case for constant $a_0$ and with this choice of $a_0$, the MOND paradigm becomes a canonical force law. Canonical MOND can also explain the Faber-Jackson relation (FJR) at large radius in spherical galaxies, Milgrom \cite{milgrom2}. Nonetheless, in other systems such as the Giant low surface density galaxies \cite{milgrom0}, spiral galaxies with bulge \cite{ angus} and  Early Type galaxies \cite{swaters} where the observed $a_0$ significantly differs from the canonical value. In dwarf galaxies, the observed MOND acceleration seems to have a very larger scatter \cite{sanders0}. In galaxy clusters, MOND acceleration $\sim 4 a_0$ is required for the central X-ray region \cite{aguirre}, whilst in the outer region the observed MOND acceleration is $\sim 20a_0$ \cite{mcgaugh2020} before it falls below $a_0$ at $2-3Mpc$ \cite{li2023}.  In the large mass Bright cluster Galaxies (BCG), the observed MOND acceleration is $7a_0-20a_0$ \cite{tian}. In large scale structure evolution simulation, the required MOND acceleration is $\sim 0.1 a_0$ to match observation, Nusser \cite{nusser2002}. Even then, at late time, the observed Newtonian acceleration dominance at solar system scale \cite{tremaine}-\cite{desmond} and  at Wide-Binary scale \cite{banik} suggests that canonical $a_0$ is too large at these scales. A possible resolution to the challenge posed by observations is to consider a MOND variant with variable $a_0$. Due to the resemblance of $a_0$ value with the cosmological acceleration $\sim cH_0$, Milgrom has considered a cosmologically varying $a_0(z)$ \cite{milgrom2015}. Other variable MOND candidate has also been considered \cite{sabine2018}. 
\\\\
Unlike dark particles, the Milgram MOND program alone does not provide the dark matter potential required to produce the CMB power spectrum \cite{mcgaugh2014} and the suppression of  Baryonic Acoustic Oscillation (BAO) in matter power spectrum \cite{dodelson}. Since the original MOND is a phenomenological theory, the hope is in finding the underlying relativistic theory from which one could obtain better understanding of the origin of $a_0$ and further insight in the mass deficit problem at large scales. These theories are usually achieved by modifying GR. Some MOND variant theories that survive the gravitational wave speed observations from GW170817 are MOG \cite{moffat} and the Relativistic MOND by Skordis \cite{skordis}. The GRB221009A observations \cite{cao} also imposes a strong constraint on the Lorentz invariance of any MOND variant model.
\\\\
Our starting point is a new metric solution of General Relativity for a point mass residing in an expanding universe \cite{wong} in which an additional MOND-like acceleration arises, where the canonical MOND acceleration $a_0$ is replaced by the cosmological acceleration $\frac{1}{2}H(z)^2r$. For Newtonian acceleration $g_N$, the effective gravitational acceleration is given by $\ddot{r}=-\sqrt{g_N a_0^{VM}}$ where $\sqrt{a_0^{VM}}=\sqrt{g_N}+\sqrt{\frac{1}{2}H^2r}$. We call $a_0^{VM}$  VMOND acceleration. This metric provides the simplest realisation of the MOND idea in a relativistic framework without modifying Einstein Gravity and therefore satisfies the observed constraints set by gravitational wave speed and Lorentz invariance. At late time $H(z)=H_0$ and solar system scale, the non-Newtonian acceleration is negligible and Newtonian acceleration dominates \cite{wong}. This avoids the problem of canonical MOND at solar system scale. 
\\\\
However, at early times the non-Newtonian acceleration can be dominant at some large scale, the obvious question is whether this new model can account for the early emergence of massive galaxy and produces the observed $a_0$ in galaxies at much later time.
\\\\
In this work we examine the late time values $a_0^{VM}$ in non-spiral galaxies such as Ellipticals and Dwarf Spheroidal galaxies. Late time Spherical galaxy is modelled as a hydrodynamical equilibrium system in which gravitational acceleration balances the pressure within the sphere and measuring its velocity dispersions at different radius will reveal the underlying gravitational acceleration. The characteristic scale is the effective (half mass) radius $r_e$. At very small distances within an effective radius of Ellipticals, Bolton et. al. \cite{bolton} obtains a confirmation of the fundamental plane relation (constant velocity disperson $\sigma_r$) which can be modelled using the "equilibrium" Jean equation of an isothermal sphere under "Newtonian" acceleration where $\sigma_r$ is constant. 
\\\\
To model the observed Velocity Dispersion-Radius relation extended well beyond $r_e$, Sanders \cite{sanders2000} introduces a small anisotropy in the form of a polytropic sphere at equilibrium, with pressure $P$, density $\rho$ and constant $K$ such that $P=K\rho^{1+\frac{1}{n}}$ with $\sigma_r^2=K\rho^{1/n}$ under the MOND acceleration. A decreasing mass density $\rho(r)$ in radial distances will then lead to a fine-tunable decrease in $\sigma_r$.  In this polytrope model, the Faber Jackson relation 
\be
\sigma_r^4=A(n,\beta) GM(r) a_0
\ee
has a structure constant $A(n,\beta)$ which is a function of $n$ and the anisotropic parameter $\beta$ and it is found that $13<n<16$ is required to fit observations. Sanders \cite{sanders2000} also highlights a degeneracy that one can also fit a data set of  mass-velocity dispersion relations of spherical galaxies in Newtonian plus Dark Matter (DM) form
\be
\sigma_r^2=B(n,\beta) \frac{GM(r)}{r_e}
\ee
where $B(n, \beta) $ includes an effective dark matter factor. This degeneracy at large radius is further investigated by  Sanders \cite{sanders2021} who tracks a run of velocity dispersions in the low mass density (deep MOND) region of low surface density spheroidal galaxies. At these large distance scales, the results are favourable towards MOND. 
\\\\
Durazo et. al. \cite{durazo}, studies the Velocity Dispersion-Radius relation of Ellpitical galaxies for a large range of mass scales and finds that these galaxies typically have a Newtonian central region characterised by an effective (half-mass) radius scale $r_{e}$ above which the velocity dispersions gradually become flat and follow a non-Newtonian Faber Jackson Relation (FJR). Similar observation for a small number of elliptical galaxies is found earlier by Milgrom and Sander \cite{milgrom4}. 
\\\\
One could try to model the evolution of an elliptical galaxy from an early time overdensity. In \cite{loeb},  it is shown that an overdensity consisting of DM and baryons, can monolithically grow from recombination to turnaround and gravitationally collapses and virialise at high redshift. Theoretical understanding of  the gravitational collapse and virialisation process of an overdensity into an equilibrium sphere is non-trivial and we will attempt to provide a more detailed recapitulation in subsection 3.2 to set-up the background for our model calculations.
\\\\
For the MOND Scheme, Sanders \cite{sanders2007} studies a spherical overdensity which evolves monolithically under MOND using simulation. It finds that small radius expanding mass shell will turnaround earlier than the large radius mass shells. This is due to an implicit comoving scale in the overdensity. Despite without a well defined MOND potential, virialisation seems plausible within a characteristic radius $r_e$ and the measured structure constant $A(n, z)$ in FJR seems compatible with values in the theoretical work \cite{sanders2000}. 
\\\\
In this work, we follow a spherical overdensity which monolithically evolves from recombination to virialisation in VMOND, assuming that the dominant relaxation process is the violent relaxation as described in the simulation works in Newtoniian gravity by van Albada \cite{vanalbada} and in MOND by Nipoti et al. \cite{nipoti}. In section 2, we recall the essential features of our model for completeness. In section 3, we consider the turnaround and collapses of the overdensity mass shells and estimate the time for the central core to reach a time independent QSS state. In Section 4, we calculate the virialised potential and estimate the late time value of $a_0^{VM}$ in the outer region for a few elliptical galaxies in different environment. Section 5 is summary and discussion.
\section{2: The model}
For a central point mass in an expanding background with Hubble parameter $H(z)$ at redshift $z$, we find a new metric that differs from the Einstein-Straus vacuole solution, having the form \cite{wong}
\begin{equation}
ds^2=c^2d\tau^2-\frac{2GMa^3}{c^2r}d\varrho^2-r^2d\Omega^2.
\label{LT1}
\end{equation}
In terms of Tolman-Lemaitre formulation, the underlying free fall velocity is given by
\be
\dot{r}= H(z)r-\sqrt{\frac{2GM}{r}},
\label{dotr1}
\ee
where $G$, $M$ and $r$ are the Newton's constant, the central point mass and the radial
distance respectively. This equation depicts that the particle will follow the Hubble law at large distances, but at small distances it will follow a Newtonian free falling velocity. Formally, Eq.(\ref{dotr1}) is similar to the radial velocity equation in Newtonian perturbation theory, which is
\begin{equation}
\dot{r}=Hr+v_p
\end{equation} 
where $v_p$ is the peculiar velocity. The radial acceleration is approximated by
\begin{equation}
\ddot{r} =\frac{\ddot{a}}{a}r+\dot{v}_p.
\end{equation}
However, in our metric the acceleration equation takes a different form
\begin{equation}
\frac{d \dot{r}}{dt}=\dot{r} \bigg(\sqrt{\frac{GM}{2r^3}}+H\bigg)+r\dot{H},
\label{ddotr0}
\end{equation}
\begin{equation}
\ddot{r}=-\frac{GM}{r^2} -\sqrt{\frac{H^2r}{2}}\sqrt{\frac{GM}{r^2}}+\frac{\ddot{a}}{a}r,\:\:\:\:\:\: \frac{\ddot{a}}{a}=-\frac{1}{2} H_m^2+\frac{c^2\Lambda}{3},
\label{nNewton1}
\end{equation}
where $H_m$ is the Hubble parameter in a matter only universe. Here we see that a non-Newtonian MOND-like acceleration arises due to the free fall relation Eq.(\ref{dotr1}). We can write the effective gravitational acceleration in Eq.(\ref{nNewton1}) without the cosmological background in the MOND form
\begin{equation}
\ddot{r} =-\sqrt{g_N a_0^{VM}},\:\:\: \sqrt{a_0^{VM}}= \sqrt{\frac{GM}{r^2} }+\sqrt{\frac{H^2r}{2}},
\end{equation}
in which for large enough $H$ and $r$, we have $\frac{1}{2} H^2r\gg g_N$. We call $a_0^{VM}$ the VMOND acceleration. The gravitational potential now takes the form 
\begin{equation}
V=\frac{GM}{r}-H\sqrt{2GMr}+\frac{1}{2}H^2r^2,
\label{phi}
\end{equation}
here the second term on the R.H.S we call the MOND-like potential.
At the turnaround distance from central mass, we have
\be
r_{ta}=\bigg(\frac{2GM}{H^2}\bigg)^{1/3}.
\ee
Although we argue elsewhere that a different choice for $H(z)$, when there are two background fluids which dominate in their own asymptotic epochs, would provide the dark matter potential required by the observed CMB angular power spectrum \cite{wong3} and matter power spectrum \cite{wong4}, as the overdensity evolution primarily occurs in the matter dominant epoch, we take the Hubble parameter at redshift $z$ to be the Friedmann-Lemaitre-Robertson-Walker (FLRW) choice
\be
H^2(z) =\frac{8 \pi G}{3}\rho_H=\frac{8\pi G}{3} \bigg(\rho_r+\rho_m+\rho_{\Lambda}\bigg)=H_0^2 \bigg(\Omega_{r}(1+z)^4+ \Omega_m(1+z)^3+\Omega_{\Lambda}\bigg),\:\:\:H^2_0=\frac{8 \pi G}{3}\rho_c,
\label{H22}
\ee
where for $i=r, m, \Lambda$, $\rho_i$ are cosmological background densities of radiation, pressureless matter which we assume is essentially baryonic matter and dark energy respectively. $H_0$  and $\rho_c$ are the Hubble parameter and the critical density of the present epoch. For $\rho_{i,0}$ denotes the density of $i$ compoent at the present epoch, we have the density parameters $\Omega_i=\frac{\rho_{i,0}}{\rho_c}$. Also the acceleration equation takes the form
\be
\frac{\ddot{a}}{a}=-\frac{4 \pi G}{3} \bigg(2\rho_r+\rho_m\bigg)+\frac{8 \pi G}{3}\rho_{\Lambda}.
\label{nNewton2}
\ee
For a local mass cloud in an expanding background, its central baryonic density $\rho_b(r)$ is given in terms of the overdensity $\delta(r)$ as
\be
\delta(r)\rho_m=\rho_b(r).
\ee
In matter dominant epoch, Eq.(\ref{nNewton1}) becomes
\begin{equation}
\ddot{r} =-\bigg(\delta +\sqrt{\delta}+1\bigg) \frac{4\pi G}{3} \rho_m r.
\label{nNewton3}
\end{equation}
In comoving coordinates, one can show that \cite{wong}, the evolution equation for $\delta$ in Newtonian gravity is replaced by $\Delta(\delta) =\delta +\delta^{1/2}$ such that the overdensity evolution equation becomes
\be
\ddot{\Delta}+2H \dot{\Delta}=4\pi G\Delta\rho_m,
\label{Delta}
\ee
which has the growth solution $\Delta =\delta+\delta^{1/2} \propto a(t)$.
Therefore at small $\delta\ll1$, $\delta^{1/2}$ dominates the function $\Delta(\delta)$, so that $\delta$ grows as $\delta \propto a(t)^2$ which is similar to the effect of MOND. As $\delta \rightarrow 1$,  the overdensity begins to grow more slowly as $\delta \propto a(t)$. 
\subsection{3.1: Over-density evolution from VMOND}
The question therefore is whether this faster overdensity evolution in our paradigm is enough to give agreement with data. We shall suggest that it can be, but the argument is not straightforward. We have to make assumptions about galaxy formation, power spectra, simulations  and energy redistribution that take us a long way beyond our original model, although they depend crucially upon its details.  
\\\\
From an initial overdensity $\delta_{int}$ at recombination, we can use $\delta+\delta^{1/2}\propto a(t)$ to calculate the overdensity $\delta$ at redshift $z<1080$ by
\begin{equation}
\delta+\sqrt{\delta}=\bigg(\delta_{int}+\sqrt{\delta_{int}}\bigg) \bigg(\frac{1081} {1+z}\bigg)=\frac{A_0}{1+z},
\label{dpsd}
\end{equation}
where $A_0$ is fixed by the value of $\delta_{int}$. At recombination, baryonic overdensity is suppressed at the Silk damping scale (a few $Mpc$). Galaxy formation is favoured near the shell's origin and at a radius of $150Mpc$ from the shell's origin, and galaxy overdensity scales are typically less than $2Mpc$.  In \cite{einasto1}, the galaxy power spectrum is found to be $\delta_k^2 \propto k^{n}$ with $n>-3$ at large $k$. In a cartesian grid approximation,  one can obtain the variance of fluctuations of smoothed density field at characteristic scale $R$ is $\delta^2 (R) \propto R^{-(n+3)}$\cite{binney} (pp720-721).  Separately, Nusser \cite{nusser2005}, uses a power law overdensity $\delta (R) \propto \delta  R^{-S},$ with $0<S<3$ for uniform $\delta$, to simulate overdensity growth under MOND. Therefore, we expect the overdensity to have an increasing power-law behaviour towards small length scales. In an uniform density approximation, the initial overdensity $\delta_{int} =\sum_R \delta(R)$ for a galaxy, should have value much higher than the $150Mpc$ average  of $10^{-5}$.
\\\\
The baseline CMB average temperature variation $\frac{\delta T}{T}=1\times 10^{-5}$ corresponds to an initial baryon overdensity $\delta_{int} =3 \times 10^{-5}$ and $A_0=5.94$. To account for large galaxies at high redshift, one needs an overdensity to turnaround at sufficiently higher redshift.
In \cite{sanders2007}, in a spherical galaxy formation under MOND, for galaxy mass $10^{11}M_{\odot}$ Sanders takes $\delta_{int} =1.8\times 10^{-3}$ which corresponds to $A_0=47.8$. (To better illustrate the effect of Eq.(\ref{dpsd}), we choose a slightly higher value $\delta_{int}=2.8\times 10^{-3}$ ($A_0=60$), although Sander's choice is sufficient for most of our purposes. This choice is also similar to the initial overdensity choice  made in \cite{loeb} in a dark matter plus baryon direct collapse model to explain the formation of a high redshift Supermassive black hole with mass $O(10^7)M_{\odot}$). 
\\\\
Given $\delta_{int}=2.8\times 10^{-3}$, we obtain the turnaround redshift $z_{ta}$ where $\delta=1$ from Eq.(\ref{dpsd}),
\begin{equation}
1+z_{ta}=\frac{A_0}{2}=30, \:\:\: z_{ta}=29.
\label{ta}
\end{equation}
We stress that  this turnaround redshift (resulting from this choice of $\delta_{int}$) is well within an observationally viable  redshift range $15\:\lsim z\: \lsim\: 50$ \cite{barkana}, from 21cm radiation.
\\\\
When compared to the Newtonian only evolution where $\Delta=\delta$, we note that
\begin{equation}
1+z_{ta}= 1081\delta_{int} =3.02, \:\:\:z_{ta}=2.02.
\end{equation}
We can see that the VMOND potential provides a significant lift of $z_{ta}$ to a much higher redshift, which could leave sufficient time for an overdensity to evolve into the  observed massive galaxies at high redshifts. 
\\\\
From Eq.(\ref{ta}), we can estimate the turnaround time when the overdensity reaches $\delta=1$.
In a matter only universe, we have the scale factor $a(t) \propto t^{2/3}$, so that
\begin{equation}
H(z)=\frac{\dot{a}}{a} =\frac{2}{3t}.
\label{mt1}
\end{equation}
At redshift $z_{ta}$ the turnaound time   is
\begin{equation}
t_{ta}=\frac{2}{3H(z_{ta})}.
\end{equation}
\subsection{ 3.2: An overdensity Journey time after turnaround}
In this subsection, we recapitulate the mechanism of direct gravitational collapse after an overdensity turnarounds in the Newtonian theory, which will provide for a basis of our calculations in the non-Newtonian VMOND model.
\\\\
An idealised galaxy cluster is described by a virialised central region of galaxies and gas at equilibrium, which can be modelled by the Jeans equation in which the central potential is Newtonian and the cosmological background potential becomes a separate entity having no effect on the central cloud. This procedure is without formal justification but its usefulness is a puzzle called "Jeans Swindle". For a more realistic cluster, outside the central virialised region there is an infalling region in which galaxies are still coming towards the centre and thus not virialised. For this type of cluster, Falco et al. \cite{falco2013} argues for a generalised Jeans equation in which the cosmological background acceleration is considered weak and negligible at radius inside the infalling region. At radius "sufficiently far outside the infalling region",  the cosmological acceleration dominates. Within the infalling region, one obtains the original Jeans equation and additional radial peculiar velocity terms which are applicable to the infalling (and unvirialised) galaxies. This explanation suggests that the Jeans Swindle can be effective due to the dominance of the Newtonian central potential inside an infalling region. 
\\\\
In the context of Newtonian perturbation theory, after the overdensity mass shells turnaround, they are assumed to decouple from the cosmological background \cite{bosch},  again assuming Jeans Swindle. Individual particle takes up its own energy and potential with $E=\frac{1}{2}v^2+\Phi<0$, the collection of these energies and potentials becomes the initial state for a graviational collapse. In Sanders' simulated gravitational collapse of an overdensity under the MOND acceleration \cite{sanders2007}, Jeans Swindle is also assumed at the outset.
\\\\
After turnaround, see Binney-Tremaine (BT) \cite{binney} (pp-379-382), the evolution of particles on the mass shells are described by the (Newtonian) Poisson equation and the collisionless Boltzmann equation which is a continuity equation for a particle's probability distribution function in phase space.  Here the density of phase points in a small volume around the phase point of any oscillating particle is assumed constant. Different initial energies of a particle will stretch the original phase space patch into a filament. When the filaments are sufficiently (phase-)mixed the resulting macroscopic distribution function will tend to its equilibrium state. 
\\\\
In a collisionless gravitational collapse, the energy loss due to physical reasons is assumed negligible, so that the total energy of the system is approximately conserved.  
Before the system comes toward equilibrium,  a particle's energy follows the equation
\begin{equation}
\frac{dE}{dt}=\frac{\partial E}{\partial v} \frac{dv}{dt} +\frac{\partial E}{\partial \Phi} \frac{d\Phi}{dt}= -v \cdot \nabla \Phi+\frac{d\Phi}{dt}=\frac{\partial \Phi}{\partial t},
\label{dedt}
\end{equation}
which means that a particle's energy will change if its potential changes in time. Based on this scenario, Lynden-Bell (LB) \cite{lyddenbell} argues that a star's potential should vary (violently) in space and time due the movements of particles during the gravitational collapse. In practice, there will be a large scale potential fluctuations which lead to a redistribution of energies between stars and provides a much faster relaxation, called "Violent relaxation". More specifically, LB
argues that in this case, the solution of the system consisting of collisionless-Boltzmann equations  involves intermingled filaments on smaller and smaller scales. For macroscopic observables, LB introducees a coarse-grained distribution function which is a local average of distribution function over the filaments. This coarse-grained distribution is expected to reach maximum entropy on a short dynamical time $t_D$. In the case of stellar systems, this "coarse-grained" distribution will tend to the Maxwell-Boltzmann distribution, but the fine-grained distribution could retain its initial distribution.
The collapse via phase mixing and violent relaxation, is called virialisation (\cite{binney} pp734). After the process is complete, in a spherical collapse, the averaged kinetic and potential energy in final sphere should satisfy the Virial Theorem, which we will use in the next section. Numerical simulations support the Violent relaxation scenario under both Newtonian gravity \cite{vanalbada} and MOND \cite{nipoti}. Violent relaxation is also assumed to be the dominant relaxation mechanism for the direct gravitational collapse model of a dark matter halo to form very high redshift supermassive blackhole \cite{loeb}. 
\\\\
In Newtonian gravity, the violent relaxation time is taken to be  the Newtonian free fall time $t_{ff}$ \cite{lyddenbell}, which depends on its initial matter density. We note that different authors use different definitions of dynamical time unit to do simulations or for analytic estimates. For example, in BT \cite{binney} pp.268, the (Newtonian) free fall time 
\begin{equation}
t_{ff}=\frac{1}{4} \sqrt{\frac{3 \pi}{2G \rho_m} }
\end{equation}
is used. Lynden-Bell \cite{lyddenbell} uses the dynamical time 
\begin{equation}
t_D=\frac{3}{4}\sqrt{\frac{1}{2\pi G \rho_m}}.
\end{equation}
In this work, we follow the definition in Nipoti et al. \cite{nipoti}, \cite{ciotti}, 
\begin{equation}
t_{dyn}=2t_N,\:\:\:t_N= \sqrt{\frac{r^3}{GM}},
\label{tdyn0}
\end{equation}
mainly because they simulate the spherical gravitational collapse of both Newtonian and MOND system, so that it provides a simpler reference time-scale for collapse in VMOND, which should be at the same order.
\\\\
In LB \cite{lyddenbell}, the estimated violent relaxation time is $t_D \sim 0.91 t_N$, therefore, a coarse-grain equilibrium central sphere can appear at $t\sim t_{ta}+ 0.91t_N$. In  the Newtonian simulation in \cite{binney} (pp 383), 
after $t=t_{ta}+1.61 t_N$, a tight minimum configuration has developed. By $t=t_{ta}+ 4.8 t_N$, the core is settled to a Quasi-Stationary State (QSS) (a not fully equilibrium but time independent state). This is attributed to the stretching and folding of filaments \cite{debuyl}. At $t=t_{ta}+9 t_N$, an equilibrium sphere is developed. However, it does not preclude the possibility that some outer large mass shells remain outside of the virialised sphere.
\\\\
This is a common feature seen in gravitational collapse simulation works, \cite{sanders2007}, \cite {vanalbada}-\cite{nipoti}, \cite{ciotti}. In general, a tight central region can form after $t=t_{ta}+t_N$. A QSS  can be reached at $t=t_{ta}+ t_{ff}+2t_N$ \cite{ciotti}, and a complete virialisation is reached at $t=t_{ta}+t_{ff}+6 t_N$, \cite{nipoti}. The gravitational collapse simulations in \cite{nipoti} and \cite{ciotti} are done for both Newtonian and MOND acceleration. 
\subsection{3.3: The effect of angular momentum on an overdensity}
In our VMOND model,  consider the acceleration equation Eq.(\ref{nNewton1}) and equivalently Eq.(\ref{nNewton3}) for an overdensity in physical coordinates.
If during overdensity growth, from tidal effects the overdense mass shells at $r$ receive a specific angular momentum $h$ such that
\begin{equation}
\frac{h^2}{r^3} = \bigg(\delta +\sqrt{\delta}\bigg)\frac{4\pi G}{3} \rho_mr,
\end{equation}
then Eq.(\ref{nNewton3}) becomes
\begin{equation}
\ddot{r} = \frac{\ddot{a}}{a}r,
\end{equation}
which simply describes the acceleration of the cosmological background.
A consistent scenario is that the overdensity stops growing and separates from the expanding background.
A particle on the overdensity mass shell will then follow its own equation of motion
\begin{equation}
E =\frac{h^2}{2r^2}+\frac{1}{2}\dot{r}^2-\frac{GM}{r}+H\sqrt{2GMr}.
\label{fpee0}
\end{equation}
where $H$ is now fixed at the value when the breakaway occurs. Here we postulate that the introduction of siginificant specific angular momentum to a particle in an overdensity is the main reason for the Jeans Swindle phenomenon, since in a FLRW expanding universe, a particle can have a peculiar velocity in the Hubble flow, but the specific angular momentum is defined around a central mass and is privy to a gravitationally bound system.
\\\\
Consider the more relevant scenario that the overdensity reaches $\delta\geq 1$ before sufficient specific angular momentum and negative particle energy $E<0$ are attained to separate the overdense region from the cosmological background. A particle on an overdensity mass shell can start infalling following Eq.(\ref{fpee0}) with $H\sim H(z_{ta})$. 
\subsection {3.4: In the VMOND model, can a Stable Central Region form before $z>6.5$ ?}
To estimate the free-fall time for a particle at turnaround radius following Eq.(\ref{fpee0}), we consider that $H=H(z_{ta})$ is fixed for $E=h=0$. Taking
\begin{equation}
y=\frac{2}{3} \sqrt{\frac{r^3}{2GM}},
\end{equation}
Eq.(\ref{fpee0}) becomes
\begin{equation}
\dot{y}=(1-3Hy)^{1/2},\:\:\:\:\:  t=\frac{2}{3H}\sqrt{1-3Hy}.
\end{equation}
At turnaround where $\dot{y}=0$ and $1-3Hy=0$, we have $t=0$.  At $y \rightarrow 0$, the free-fall time $t_{ff}= \frac{2}{3H(z_{ta})} =t_{ta}$. 
\\\\
If we assume that the fine-grained Collisionless Boltzmann equation and Poisson equation retain nearly its initial values, the $H(z)$ appeared in the virialised potential will remain at $H=H(z_{ta})$.
\\\\
In \cite{ferreira2}, the image of a large and bright central region is given for the very high redshift $z>11$ galaxies, Furthermore the galaxy morphologies is given for the $1.5<z<6.5$. However, it does not provide enough information on how far the galaxy virialisation has progressed. (We leave the consideration of the observed supermassive blackholes and Little Red Dots at very high redshifts $z\: \gsim\: 6$ in another work.) The tight central configuration obtained after $t_{ff}$ could arguably provide a reasonable description of the observed bright central region in high redshift galaxies.
\\\\
At $\delta=1$, the overdensity mass inside $r_{ta}$ is given by $M=\frac{4\pi}{3}\rho(z_{ta}) r_{ta}^3$, so that from Eq.(\ref{tdyn0}) $t_{dyn}$ takes the value
\begin{equation}
t_{dyn}=2\sqrt{\frac{r_{ta}^3} {GM}}=\frac{2\sqrt{2}}{H(z_{ta})}=2t_N
\end{equation}
The time to virialisation $t_{vr}$ is given by
\begin{equation}
t_{vr}=t_{ta}+t_{ff}+ 2Nt_{N} =\frac{2}{3H(z_{ta})} \bigg(2+3\sqrt{2}N\bigg)=\frac{2}{3H(z_{vr})},
\label{tvr}
\end{equation}
where $z_{vr}$ is the virialisation redshift and $N$ is the number of $t_{dyn}=2t_N$ after $t_{ff}$ required to reach different level of virialisation. From Eq.(\ref{tvr}), within matter dominant epoch we obtain
\begin{equation}
1+z_{vr}=\frac{1+z_{ta}}{(2+3\sqrt{2}N)^{2/3}}.
\label{zvrzta}
\end{equation}
Since in VMOND, the $t_{ff} =t_{ta}=0.47t_N$ which is shorter than $t_D \sim 0.91 t_N$ in the Newtonian violent relaxation \cite{lyddenbell}. The faster free fall time implies that the mixing would be more violent and we assume that after $t_{ff}$ ($N=0$), a tight minimum configuration could occur. Eq.(\ref{zvrzta}) shows that this occurs at $z_{vr}=17.9$.
\\\\
For a Quasi-Stationary-State (QSS) obtained in simulation \cite{binney}, \cite{ciotti}, on taking $N=1$, QSS occurs at
\begin{equation}
z_{vr}=\frac{1+z_{ta}}{3.39}-1=7.85.
\end{equation}
An effectively complete virialisation occurs at $N=3$ \cite{nipoti}, where
\begin{equation}
z_{vr}=\frac{1+z_{ta}}{6.0}-1=3.99.
\end{equation}
The turnaround redshift $z_{ta} =29$ from our initial overdensity is within the Barkana-Loeb constraint \cite{barkana}. The analysis above suggests that with our chosen $\delta_{int}$, the VMOND potential is sufficient to provide for a tight minimum galactic configuration at redshift $z_{vr}\sim 17.9$, and a stable (Quasi-Stationary State) galactic core by $z\geq 7$. This central core will remain stable while evolving through the range $1.5<z<6.5$, as observed  in \cite{ferreira2}. 
 %
 %
\section{4.1: The velocity disperson in a virialised VMOND potential }
We consider a protogalactic cloud that has gravitationally collapsed and virialised under the VMOND acceleration. Given the time average kinetic energy $T=\frac{1}{2}M \bar{v}^2$  with the average potential $V$, the virialised relation is $2T+V=0$. The system's total energy $E_T$ is given by $E_T=T+V$. Since the total energy remains unchanged and the average velocity $\bar{v}$ does not change much over the relaxation, the average potential $V$ should retain the value of the average potential at the system's initial state. 
\\\\
Next we calculate the virialised potential of the system.  We work with the mass density of the virialised cloud at distance $r$ given by 
\begin{equation}
\rho_m(r) =\bar{\rho} r^{-S}
\label{rhom}
\end{equation}
where $\bar{\rho}$ is a constant with  $0<S<3$.
\\\\
In general, the central mass upto a shell at $r$ is
\begin{equation}
M(r)=4 \pi \int_0^{r} r^2 \rho_m(r) dr =\frac{4 \pi \bar{\rho}} {(3-S)} r^{(3-S)}.
\label{mt}
\end{equation}
The virialised potential \cite{binney} is given by
\begin{equation}
V(R)=-4\pi\int_0^{R} drr^3 \rho_m(r) \nabla \cdot \Phi .
\label{vp}
\end{equation}
For the Newtonian acceleration, using $M(r)$ in Eq.(\ref{mt}) 
\begin{eqnarray}
V(R) &=-& 4\pi \int_0^R dr r^3 \bar{\rho} r^{-S} \frac{GM(r)}{r^2}=-G\frac{(4\pi \bar{\rho})^2}{3-S}\int_0^R r^{4-2S} dr
\nonumber
\\
&=& -G\frac{(4\pi \bar{\rho})^2}{(3-S)} \frac{R^{5-2S} }{(5-2S)}=-c_1(S) \frac{GM^2}{R};\:\:\:c_1(S) =\bigg(\frac{3-S}{5-2S}\bigg).
\end{eqnarray}
where $M=M(R)=4\pi \bar{\rho} R^{3-S}/(3-S)$.
For an uniform density, $S=0$, one recovers the virial potential $V=-\frac{3}{5}\frac{GM^2}{R}$  (BT \cite{binney}, pp 64).
For $S=2$, one obtains the Newtonian potential energy $V=-\frac{GM^2}{R}$.
The calculation above suggests that the virialised potential Eq.(\ref{vp}) provides a reasonable description of the effective potential of the virialised object.
\\\\
For the VMOND acceleration, assuming for the time being that there is "no potential energy loss" during virialisation, the virialised potential is given by
\begin{eqnarray}
\Delta V(R)=-\frac{H(z_{ta})}{\sqrt{2}} 4\pi \bar{\rho} \sqrt{\frac{4\pi G\bar{\rho}} {(3-S)}}\int_0^R dr r^{5/2-S} r^{3/2-S/2},
\end{eqnarray} 
\begin{eqnarray}
\Delta V(R) =-\frac{H(z_{ta})}{\sqrt{2}} 4\pi \bar{\rho} \sqrt{\frac{4\pi G\bar{\rho}} {(3-S)}} \frac{R^{5-\frac{3}{2}S}}{5-\frac{3}{2}S}
\end{eqnarray} 
That is,
\begin{equation}
\Delta V(R) =-M H(z_{ta}) \sqrt{\frac{GMR} {2}} c_2(S),\:\:\:\: c_2(S)=\bigg(\frac{3-S}{5-\frac{3}{2}S}\bigg).
\label{nonNewptl}
\end{equation}
For a stationary orbit, the kinetic energy is dominated by an averaged $3-D$ velocity dispersion $v^2=\sigma^2=3\sigma_r^2$, where $\sigma_r$ is the velocity dispersion in the radial direction. The virial theorem becomes 
\begin{equation}
\sigma^2=c_1(S)\frac{GM}{R}\bigg(1+\frac{c_2(S)}{c_1(S)}\sqrt{\frac{H^2(z_{ta})R^3}{2GM(R)}}\bigg) .
\label{virialeq1}
\end{equation}
With an emphasis on the long distance non-Newtonian behaviour,  we obtain
\begin{equation}
\sigma^4 =c_2(S)^2 GM\bigg(\frac{1}{2} H^2(z_{ta}) R\bigg) \bigg(1+\frac{c_1(S)}{c_2(S)} \sqrt{ \frac{2GM}{H^2(z_{ta}) R^3}}\bigg)^2.
\label{sigmara0}
\end{equation}
From the observations in \cite{durazo}, a more realistic mass density profile is described by 
\begin{equation}
\rho_m(r) =\rho_0r^{-S}+\rho_{\infty},
\label{rhom}
\end{equation}
where $\rho_0$ and $\rho_{\infty}$ are constant with  $S>0$.  Because of the different galactic sizes discussed in \cite{durazo}, we introduce a reference scale $r_{ta}$ such that $n=\frac{r_{ta}}{R}$. At small distances where the $\rho_0$ term dominates, we have
\begin{equation}
\sigma^2=c_1(S) \frac{GM}{R}\bigg(1+\frac{c_2(S)}{c_1(S)}\sqrt{\frac{3-S }{3} \frac{1}{n^S}}\bigg).
\label{virialeq02}
\end{equation}
At long distances where $\rho_{\infty}$ term dominates, the enclosed mass inside radius $R$ is nearly constant at $M=\frac{4\pi}{3}\rho_m r_{ta}^3$,  we obtain
\begin{equation}
\sigma^4= c_2(S)^2GM \bigg(\frac{1}{2}H(z_{ta})^2R\bigg) \bigg(1+\frac{c_1(S)}{c_2(S)}n^{3/2}\bigg)^2.
\label{vsigma02}
\end{equation}
\subsection{ 4.2: The velocity dispersion - $a_0$ relation}
In elliptical galaxy, the observables are the velocity dispersions $\sigma$ and central mass $M$  (through a best estimated mass to light ratio). The observed mass deficit in the $\sigma$-$M$ relation in Newtonian gravity at large distances is attributed to a dark particle halo or a new gravitational acceleration scale $a_0$. In MOND, from the observed $\sigma$ and $M$, the value of $a_0$ is obtained with an assumed density profile and MOND interpolation function. For example, Milgrom \cite{milgrom} uses a singular isothermal sphere ($\rho(r) \propto r^{-2}$) to evaluate the large radius limit of 3-D isotropic velocity $\sigma$. The MOND acceleration $a_0$ is given by
\begin{equation}
\sigma^4=\frac{4}{9}GMa_0,
\label{mila0}
\end{equation}
However, in a study involving spherical galaxies over 7 order of magnitude in mass \cite{durazo}, where more general density profile is appropriate. The following $\sigma-M$ relation is used.
\begin{equation}
\sigma^4 =\frac{1}{9}GMa_0.
\label{durazoa0}
\end{equation}
\subsection{4.3: The $\sigma(R)$ profile of spherical galaxies}
For elliptical galaxies, the characteristic scale is the half mass radius $R_e$ (the radius from within which half of the galaxy mass is contained). \cite{durazo} considers spherical galaxies over 7 orders of magnitude in mass, it is found that for some galaxies the velocity dispersion is nearly constant for $r\ll R_e$. This constant velocity dispersions profile inside $R_e$ can be modelled by Newtonian gravity with density $\rho(r)\propto r^{-2}$ which leads to $M(R)\propto R$. The small $R$ velocity dispersion is well described by the large $n$ Newtonian potential in Eq.(\ref{virialeq02}) such that $\sigma^2\sim \frac{GM(R)}{R} =constant$.
\\\\
As $R$ increases, \cite{durazo} presents the $\sigma(R)$ profile for different elliptical galaxies without providing individual galactic masses. For NGC 5139 and NGC 0677, more careful profiles are given for the ratio $\sigma(R)/\sigma_{\infty}$ in Fig.3 \cite{durazo}, in which $\sigma_{\infty}$ is the large radius asymptotic value. $\sigma(R)/\sigma_{\infty}$ initially decreases from $R\ll R_{\sigma}$ non-linearly and transitions at $R_{\sigma}$ to a nearly constant profile at large $R$. For NGC 5139, this $R_{\sigma}=16.7pc$ which is slightly larger than the half mass radius $R_e=10.42pc$. $\sigma(R)/\sigma_{\infty}$ enters completely into the flat region at $R=2R_{\sigma} \sim 3 R_e$.
\\\\
To see whether we can obtain the profile in Fig. 3 \cite{durazo} using Eq.(\ref{virialeq02})  and Eq.(\ref{vsigma02}). For $n=\frac{r_{ta}}{R}$, we consider the range of $R$ such that $130>n>1$. 
\\\\
To start, we consider the large $R$ ($n\rightarrow 1$) region,  we assume that the effective density is $\rho_{\infty}$ and choose $S=0$ (so that $c_1(S)=c_2(S)=3/5$), whence Eq.(\ref{vsigma02}) becomes
\begin{equation}
\sigma^4= GM\bigg(\frac{3}{5} \bigg)^2\bigg(\frac{1}{2} H_0^2 \bigg)\Omega_b (1+z_{ta})^3\frac{r_{ta}}{n}\bigg(1+n^{3/2}\bigg)^2.
\label{sigmar}
\end{equation}
\begin{equation}
\sigma_{\infty}^4 =4GM( \bigg(\frac{3}{5}\bigg)^2\bigg(\frac{1}{2} H_0^2 \bigg)\Omega_b (1+z_{ta})^3r_{ta}.
\end{equation}
\begin{equation}
\frac{\sigma (R)}{\sigma_{\infty}} =\bigg[\frac{1}{4n}  \bigg(1+n^{3/2}\bigg)^2\bigg]^{1/4} \rightarrow 1\: (n \rightarrow 1). 
\label{sigmaratio}
\end{equation}
For small $R<r_{ta}$, $n=\frac{r_{ta}}{R} \gg 1$ region, asymptotically for large $n$  we have $S=2$ and $C_1(S) =1$ and Eq.( \ref{virialeq02}) becomes
\begin{equation}
\sigma^2 = \frac{ GM}{R} \bigg(1+\frac{1}{2n}\bigg).
\end{equation} 
In the large n region, as a first approximation we can neglect the $1/2n$ term but keeping a general $S$ in its density profile and $c_1(S)$. Using the definition of $M(R)$ as defined in Eq.(\ref{virialeq1}), we obtain
\begin{equation}
\sigma^2= \bigg(\frac{ 4\pi \rho_0  r_{ta}^{2-S}}{5-2S}\bigg)  n^{S-2}.
\end{equation}
We notice that for $S=2$, $\sigma(R)$ is a constant and one can normalise $\sigma(R)/\sigma_{\infty} \sim 2.4$ at $R\rightarrow 0$ as described in Fig. 3 of \cite{durazo}. To account for the decreasing profile of $\sigma(R)/\sigma_{\infty}$ for increasing $R$, we set $S=2+\epsilon(R)$ so that 
\begin{equation}
\frac{\sigma(R)}{\sigma_{\infty}} \propto \frac{1}{1-2\epsilon(R)} n^{\epsilon(R)}.
\end{equation}
We need $0<\epsilon(R)\ll1$ for decreasing $n$ (increasing $R$) to obtain qualitatively the decreasing $\sigma(R)/\sigma_{\infty}$ profile until the transition radius $R\sim R_{\sigma}$ where the non-Newtonian acceleration begins to be important. From Eq.(\ref{sigmaratio}), we assume that the galactic density at larger radius (eg $R\sim 6R_e$) can be modelled by an effectively uniform density $\rho_{\infty}$ and the enclosed mass around $6R_e$ is the total galaxy mass and is not sensitive to radius change.  To further test the validity of Eq.(\ref{vsigma02}), we will calculate the $a_0^{VM}$ value at large distances where the non-Newtonian acceleration dominates and compare with observations.
\subsection{4.3 Value of $a_0^{VM}$ for large mass galaxy}
From the above discussions we know that $\sigma(R)$ will enter into a flat region at large $R$. Using Eq.(\ref{vsigma02}) and the constraint Eq.(\ref{durazoa0}), one can calculate the large $R$ value of $a_0^{VM}$ which is not {\it a priori} fixed to a canonical value.
For this purpose, we choose, from \cite{suess}, a large star forming elliptical galaxy with mass $M_{*}=10^{10.5} M_{\odot}$ with $R_e\sim 3kpc$ and mass data is given within a radius $r_{18}=6R_e$. We use the Milky way "gas to star" ratio $M_g=(3.43/6.07) M_{*}$ \cite{wong2}.  From $M=M_{*}+M_g$ and $\rho(M)=\rho(z_{ta})$ at $\delta=1$, 
However, for the present time critical density $\rho_c$, the spherical mass $M$ is fixed at $z_{ta}$ by
\begin{equation}
 M=\frac{4\pi}{3}\Omega_b \rho_c (1+z_{ta})^3r_{ta}^3,
\label{Mzta}
\end{equation}
where $\Omega_b$ is the baryon density parameter. We find that $r_{ta}=39.2kpc$.
To evaluate $a_0^{VM}$, we make a simplification by writing phenomenological MOND acceleration $a_0$ in multiples of $\frac{1}{2}H_0^2 r_{18}$, where $r_{18}=6R_e$, the factor $n =\frac{r_{ta}}{6R_e}=2.18$.
\begin{equation}
a_0=\frac{1}{6}cH_0=\bigg(\frac{1}{2}H_0^2r_{18}\bigg)\gamma;\:\:\gamma=\frac{c/H_0}{3\times 18kpc}=7.41\times 10^4,
\end{equation}
From  Eq.(\ref{durazoa0})
\begin{equation}
a_0^{VM}= \frac{9\sigma^4}{GM}=\bigg(\frac{1}{2}H_0^2 r_{18} \bigg) \gamma.
\end{equation}
Choosing $z_{ta}=29$ and $\Omega_b=0.05$, (assuming $S=0$ for large radius) our modelled $\gamma$ is given by
\begin{equation}
\gamma= 9\bigg(\frac{3}{5}\bigg)^2 \Omega_b (1+z_{ta})^3 \bigg(1+n^{3/2}\bigg)^2. 
\end{equation}
We obtain that
\begin{equation}
\gamma=7.78 \times 10^4,\:\:\: a_0^{VM}(6R_e)=1.05 a_0,
\end{equation} 
in remarkable qualitative agreement with the $a_0$ of MOND asnalysis. In other cases, \cite{durazo} and \cite{milsanders}, the velocity dispersions can be observed upto to $R=7R_e$. At this distances, the same calculation yields $a_0^{VM}(7R_e)=0.86a_0$. 
\\\\
Next, we consider a larger galaxy at the scale of BCG with mass $10^{11.5}M_{\odot}$, which has small $R_e =1-4kpc$. If we assume $z_{ta}=29$ as before, the turnaround radius for the more massive galaxy is larger at $r_{ta}=2.15\times 39 kpc$. We can evaluate $a_0^{VM}$ around $15kpc$ and $18kpc$ as before and obtain
\begin{equation}
a_0^{VM}(15kpc) =9.94a_0,\:\:\:\:\: a_0^{VM}(18kpc)=7.37a_0.
\end{equation}
In \cite{tian}, for 54 MaNGA BCGs the observed MOND acceleration is $a_0^{VM}=7.5a_0-20a_0$, which match well with our theoretical predictions.
\\\\
In the VMOND model, the fact that from a realistic overdensity at recombination that we can obtain the canonical MOND acceleration ($a_0^{VM}\sim a_0$) for galactic mass  $10^{10.5}M_{\odot}$ and the observed non-canonical MOND value ($a_0^{VM}\sim 10 a_0$) for galactic mass $10^{11.5}M_{\odot}$ is clearly an important result which implicates that the MOND acceleration $a_0$ is a derived scale, not a fundamental constant.
\subsection{4.4: A Brief Note on Dwarf Spherical Galaxy}
Kinematics of Tidal Dwarf Galaxies (TDG) which are formed by debris at late time $z\simeq 0$ is found to be compatible with Newtonian gravity with no Dark matter \cite{lelli3}. It also poses a challenge to the canonical MOND paradigm. In the VMOND paradigm, the turnaround redshift $z_{ta} \sim 0$, the non-Newtonain potential in Eq.(\ref{virialeq1}) is at $O(10^{-3})$ times the Newtonian potential so that the velocity dispersion should be dominated by Newtonian acceleration dynamics.
\\\\
On the other hand, low surface brightness spheroidal galaxies around the Milky Way galaxy are found to have large velocity dispersions and for some galaxies the velocity dispersions are compatible with MOND \cite{sanders2021}.  Here we consider the galaxy Carina in \cite{sanders2021} where $M=2.4\times 10^6M_{\odot}$, the effective radius $r_{eff}=0.369\: kpc$, the line of sight velocity dispersions is $6.6\pm 2.6\: km/s$. We assume that Carina is a small part of the large overdensity which formed the Milky Way. Instead of gravitationally collapsing to the Milky Way mass centre, this small cloud also decouples from the cosmic background at $z_{ta}=29$ at sufficiently high systematic angular momentum which keeps the small gas cloud to stay in a large stable (rotating) orbit around the Milky Way centre. This Dwarf galactic gas subsequently gravitationally collapses onto its own mass centre which lies in the large rotating orbit. We assume that this small cloud's gravitational collapse to its mass centre and the virialise process would follow the collapse and virialisation process discussed in subsection 3.2 above, such that both Newtonian and non-Newtonian acceleration are present in the virialised dwarf galaxy. For the mass of Carina, the turnaround radiius is $2.67 kpc$. Consider $R=0.5kpc$, $n=5.34$
\begin{equation}
\gamma=9\bigg(\frac{3}{5}\bigg)^2 \bigg(\Omega_b (1+z_{ta})^3 \bigg (\frac{0.5kpc}{18kpc}\bigg)\bigg) (1+n^{3/2})^2 = 2.16\times 10^4.
\end{equation}
The velocity dispersion 
\begin{equation}
\sigma(0.5kpc)=\bigg(\frac{1}{9} GM a_0^{VM}\bigg)^{1/4}= 5.9 km/s.
\end{equation}
This is within the line of sight observation scatter $\sigma = 6.6\pm 2.6\: km/s$ in \cite{sanders2021}.
\section{5: Summary and Discussion}
$\Lambda CDM$ model is challenged by Hubble tension, the observed large mass galaxies and supermassive blackholes at very high redshift and canonical MOND is facing observational difficulties at late time solar system scale.
For a central mass in an expanding background, our proposed solution is a new metric to General Relativity in which a variable MOND acceleration arises in its geodesic equation. This MOND variant solution comes within a relativistic framework and therefore avoids the constraints set by Gravitational wave speed and Gamma Ray observations on Lorentz invariance. The non-Newtonian MOND-like acceleration, which depends on $H(z)$ and the radius, is found to fall below Newtonian gravity in our solar system and thus match observations that are problematic for the canonical MOND program \cite{wong}.
\\\\
In this work we use this new VMOND model to consider elliptical galaxy formed monolithically from a reasonable baryon overdensity at recombination. We find that the turnaround redshift is around $z_{ta}\sim 29$ instead of $z_{ta}\sim 2$ in the Newtonian perturbation model. After overdensity mass shells turnaround, assuming the dominant relaxation mechanism is the Violent relaxation which is a common assumption in both Newtonian and MOND monolithic gravitational collapse models. We borrow the gravitational collapse simulation results in the literature under Newtonian and MOND acceleration to show that, in our VMOND model, a central core can form and reach a time independent Quasi Stationary State at $z\:\geq 7$ which could explain the JWST result that a large number of galaxies can evolve through $1.5>z>6.5$ without morphology change. 
\\\\
In the virialised potential, the modelled averaged Velocity dispersons-Radius relation could match qualitatively observations in Durazo et al. \cite{durazo}.   Given a galaxy mass, we can calculate $a_0^{VM}$ using the equilibrium velocity dispersion values obtained from the galaxy virialised potential at large radius. We find that for massive galactic scale $M=10^{10.5}M_{\odot}$, $a_0^{VM}\sim a_0$ at $6R_e$ which is the typical observation radius limit. At Bright Cluster Galaxy scale  $M>10^{11.5}M_{\odot}$, we find that at $6R_e$, $a_0^{VM} > 7.3 a_0$ which is consistent with the observed non-canonical MOND value $ 7.2a_0-20a_0$ for a group of BCGs. 
Tidal Dwarf Galaxies are formed by debris at late time. The non-Newtonian potential is evaluated at $z_{ta}\sim 0$ which has a much smaller value than the Newtonian potential and the galaxy kinematics therefore should follow the dominant Newtonian acceleration (no dark matter) dynamics \cite{lelli3}. From our model, we show also that it is possible to obtain the observed high velocity dispersions in Dwarf Spheroidal galaxy orbiting around the Milky Way centre at large radius. 
\\\\
In Summary: Application of the VMOND model to spherical galaxy formation and evolution results in late time large radius $a_0^{VM}$ value which is both non-canonical and matches observations. These remarkable results show that the MOND scheme and its non-Newtonian acceleration scale $a_0$ could be the result of a more fundamental acceleration given in the VMOND metric.  The case for spiral galaxy is reported in \cite{wong2}. The missing mass problem in CMB power spectrum and Matter power spectrum observations are addressed elsewhere \cite{wong3}-\cite{wong4}.
\section*{Acknowledgements}
We thank Prof. Salucci for helpful comments on an earlier version of the text.
\section*{Data availability statement}
Data Sharing not applicable to this article as no datasets were generated or analysed during the current study.

\section*{Competing Interests}
The authors have no conflicts of interest to declare that are relevant to the content of this article.

\end{document}